\documentclass{article}

\usepackage{amssymb}
\usepackage{amsmath}
\usepackage{physics}
\usepackage{latexsym}

\usepackage{verbatim}
\usepackage{mathrsfs}
\usepackage{amsfonts}

\usepackage{graphicx}
\usepackage{subcaption}
\usepackage{epsfig}

\usepackage{units}
\usepackage{overpic}
\usepackage[utf8]{inputenc}
\usepackage{rotating}
\usepackage{enumerate}

\usepackage{soul}

\usepackage{xfrac}
\usepackage{xcolor}
\usepackage{multirow}

\usepackage[hidelinks]{hyperref}
\hypersetup{
    colorlinks,
    linkcolor={red!50!black},
    citecolor={blue!50!black},
    urlcolor={blue!80!black}
}

\usepackage{ragged2e}

\usepackage[margin=0.9in]{geometry}
\usepackage{authblk}

\begin{document}

\newcommand{\re}{\mbox{Re}}
\newcommand{\im}{\mbox{Im}}
\newcommand{\diag}{\mbox{diag}}
\newcommand{\Real}{\mathbb{R}}
\newcommand{\Complex}{\mathbb{C}}

\newcommand{\argelia}[1]{\textcolor{red}{{\bf Argelia: #1}}} 
\newcommand{\dario}[1]{\textcolor{red}{{\bf Dario: #1}}}  
\newcommand{\juanc}[1]{\textcolor{olive}{{\bf JC: #1}}}
\newcommand{\juan}[1]{\textcolor{cyan}{{\bf Juan B: #1}}}
\newcommand{\alberto}[1]{\textcolor{blue}{{\bf Alberto: #1}}}
\newcommand{\miguela}[1]{\textcolor{red}{{\bf MiguelA: #1}}}    
\newcommand{\mm}[1]{\textcolor{orange}{{\bf MM: #1}}}     
\newcommand{\OS}[1]{\textcolor{blue}{{\bf OS: #1}}}

\title{Gravitational atoms beyond the test field limit:\\ The case of Sgr~A* and ultralight dark matter}

\author[1]{Miguel Alcubierre}
\author[2]{Juan Barranco}
\author[2]{Argelia Bernal}
\author[3]{Juan Carlos Degollado}
\author[2]{Alberto Diez-Tejedor}
\author[4,\textdagger]{Miguel Megevand}
\author[1,5]{Dar\a'io N\a'u\a~nez}
\author[6,7]{Olivier Sarbach}

\affil[1]{\em Instituto de Ciencias Nucleares, Universidad Nacional
  Aut\'onoma de M\'exico, Circuito Exterior C.U., A.P. 70-543,
  Coyoac\'an, M\'exico 04510, CdMx, M\'exico}
\affil[2]{\em Departamento de Física, División de Ciencias e Ingenierías, Campus León, Universidad de Guanajuato, C.P. 37150, León, México}
\affil[3]{\em Instituto de Ciencias F\'isicas, Universidad Nacional Aut\'onoma de M\'exico,
Apdo. Postal 48-3, 62251, Cuernavaca, Morelos, M\'exico}
\affil[4]{\em Instituto de F\'isica Enrique Gaviola, CONICET. Ciudad Universitaria, 5000 C\'ordoba, Argentina}
\affil[5]{\em Departamento de Matemática da Universidade de Aveiro and Centre for Research and Development in Mathematics and Applications (CIDMA), Campus de Santiago, 3810-183 Aveiro, Portugal}
\affil[6]{\em Departamento de Matem\'aticas Aplicadas y Sistemas, Universidad
Aut\'onoma Metropolitana-Cuajimalpa (05348) Cuajimalpa de Morelos, Ciudad
de M\'exico, M\'exico}
\affil[7]{\em Instituto de F\'isica y Matem\'aticas,
Universidad Michoacana de San Nicol\'as de Hidalgo,
Edificio C-3, Ciudad Universitaria, 58040 Morelia, Michoac\'an, M\'exico}

\affil[$\dagger$]{Corresponding Author. E-mail: mfmegevand@gmail.com}

\date{\today}

\maketitle

\begin{abstract} 
We construct {\it gravitational atoms} including self-gravity, obtaining solutions of the Einstein-Klein-Gordon equations for a scalar field surrounding a non-rotating 
black hole in a quasi-stationary approximation. 
We resolve the region near the horizon as well as the far field one. 
Our results are relevant  in a wide range of masses, from ultralight to MeV  scalar fields
and for black holes ranging from primordial to supermassive. 
%For instance, a system with a scalar field consistent with ultralight dark matter and a black hole mass  comparable to that of Sagittarius~A* can be modeled. 
For instance, the core of a galactic halo consistent with ultralight dark matter and a black hole with a mass comparable to that of Sagittarius A* can be modeled.
A density {\it spike} near the event horizon, although present, is negligible,
contrasting with the prediction in
[P. Gondolo and J. Silk,  {\it Phys. Rev. Lett.}, 83:1719–1722, 1999]
for cold dark matter.
\end{abstract}

%\maketitle

%%%%%%%%%%%%%%%%%%%%%%%%%%%%%%%%%%%%%%%%%%%%%%
% Introduction
%%%%%%%%%%%%%%%%%%%%%%%%%%%%%%%%%%%%%%%%%%%%%%

%{\it Introduction.}|
\section{Introduction}
Efforts to disentangle dark matter (DM) properties using direct~\cite{PandaX-4T:2021bab,LZ:2022lsv,XENON:2023cxc} or indirect~\cite{Song:2023xdk,Chan:2019ptd,Barman:2022jdg} detection methods have reported null results, pointing towards the fact that DM may only interact through gravity.
Experimental and observational searches have disfavored heavy DM candidates, including massive compact halo objects (MACHOs) and  weakly interacting massive particles (WIMPs). 
This has led to growing interest in lighter DM candidates, often with masses below that of the proton.
Examples of these include spin zero particles such as the QCD axion~\cite{Peccei:1977hh, Abbott:1982af,Dine:1982ah,Preskill:1982cy}, for which  $10^{-6}\,\textrm{eV} < m_a c^2 <10^{-3}\,\textrm{eV}$, axion-like particles (ALPs)~\cite{Marsh:2015xka} with masses lighter than the QCD axion, $10^{-26}\,\textrm{eV} < m_{\textrm{ALP}} c^2 <10^{-6}\,\textrm{eV}$, and fuzzy DM~\cite{Hu:2000ke}, also known as ultralight DM or scalar field (SF) DM~\cite{Matos:1999et, Matos:2000ss,Hui:2016ltb}, where $m_{\textrm{SFDM}} c^2\sim 10^{-22}\,\textrm{eV}$.

For scalar light candidates, a rich phenomenology arises when their reduced Compton wavelength $\lambda_\phi=\frac{\hbar}{m_\phi c}$ ($m_\phi$ being the DM particle mass) 
becomes macroscopic. 
%\st{,  
%in contrast to heavier particles for which $\lambda_\phi$ is so small that the particle's wave nature is irrelevant.}
In previous works~\cite{Barranco:2011eyw, Barranco:2012qs,Barranco:2013rua}, we have explored configurations of a SF surrounding a non-rotating black hole (BH), showing that, in the test field approximation they can form quasi-bound states provided that $\lambda_\phi$ is larger than twice the Schwarzschild radius {\bf $R_{\rm Sch}=\frac{2GM_{BH}}{c^2}$}, i.e.
\begin{equation}
\left(\frac{4 G M_{BH}}{c^2}\right)\left(\frac{m_\phi c}{\hbar}\right)=\frac{2 R_{\rm Sch}}{\lambda_\phi}<1\,,
\label{incertidumbre}
\end{equation}
where $M_{BH}$ is the black hole mass.
Furthermore, if the left-hand side of Eq.~(\ref{incertidumbre}) is sufficiently small, this scalar cloud is not radiated away or rapidly swallowed by the BH; on the contrary, it can survive for cosmological times~\cite{Barranco:2012qs}.
Moreover, its spectrum resembles that of the hydrogen atom, $(\hbar/m_{\phi}c^2)\omega_n=\sqrt{1-\alpha_G^2/n^2}$~\cite{Detweiler:1980uk,Barranco:2012qs,Dolan:2007mj}, with $\alpha_G$ defined in Eq.~(\ref{eq.aphaG}) below.
Configurations with these characteristics have been named scalar wigs in~\cite{Barranco:2012qs,Barranco:2013rua}, and are also known as gravitational atoms in the context of superradiance~\cite{
Baumann:2019eav,Baumann:2021fkf} (see Refs.~\cite{Urena-Lopez:2002nup,Cruz-Osorio:2010nua} for related discussions).

%{\bf Autogravitante pero aproximaciones no claras ahora~\cite{Davies:2019wgi, Bar:2019pnz,Li:2020qva}. En realidad es otro modelo de materia oscura pero nos pidieron cita~\cite{DeLuca:2023laa,Berezhiani:2023vlo}} \OS{hay que quitar este parrafo!}

However, when considering SF configurations whose total mass $M_T$ is comparable to or larger than $M_{BH}$, effects resulting from the self-gravity of the SF must be included. Due to the fact that the black hole continuously accretes SF, this leads to an additional challenge: the resulting configurations cannot be exactly stationary. First attempts to include the self-gravity of the SF have been made in~\cite{Sanchis-Gual:2014ewa,Okawa:2014nda, Okawa:2015fsa, Barranco:2017aes, Hui:2019aqm,Clough:2019jpm,Cardoso:2022nzc,Pantig:2022sjb,Aurrekoetxea:2023jwk,DellaMonica:2023dcw,Aurrekoetxea:2024cqd}.\footnote{Related studies include Newtonian approximations of the problem~\cite{Davies:2019wgi, Bar:2019pnz,Li:2020qva} and the consideration of non-canonical scalar field dark matter models~\cite{DeLuca:2023laa,Berezhiani:2023vlo}.
} In particular, Ref.~\cite{Barranco:2017aes} develops a quasi-stationary approximation using Schwarzschild-type coordinates and the specification of suitable inner boundary conditions at a surface lying outside the horizon. However, this approach is limited by the fact that this surface cannot be placed arbitrarily close to the horizon, making it difficult to control the errors.

By introducing a new set of coordinates that generalize the ingoing Eddington-Finkelstein ones to the self-gravitating case, in this work we improve the approach of~\cite{Barranco:2017aes} and construct spherically symmetric solutions of the Einstein-Klein-Gordon (EKG) equations describing a SF that surrounds a BH in the quasi-stationary approximation. We shall call these configurations {\it self-gravitating gravitational atoms}. As shown in this letter, our improvement allows one to substantially extend the parameter space to include astrophysically relevant examples which were not accessible using the previous approach~\cite{Barranco:2017aes}. Furthermore, the accuracy improvements in the region near the black hole allow us to study phenomena that were not possible to find in that previous work, like the existence and characteristics of density spikes.

%\st{By introducing a new set of coordinates that generalize the ingoing Eddington-Finkelstein ones to the self-gravitating case, in this work we further develop the approach of ... and construct gravitational atoms in spherical symmetry including self-gravity, obtaining quasi-stationary solutions of the Einstein-Klein-Gordon (EKG) equations describing a SF that surrounds a BH.}

In terms of the {\it gravitational fine structure constant}~\cite{Arvanitaki:2014wva}, defined by
\begin{equation}\label{eq.aphaG}
\alpha_{G}\equiv \frac{GM_{BH}m_\phi}{\hbar c}=7.5\times 10^{9}\left(\frac{M_{BH}}{M_\odot}\right)\left(\frac{m_\phi c^2}{\unit{eV}}\right),
\end{equation}
the condition~(\ref{incertidumbre}) for the existence of SF clouds in the test field limit
becomes $\alpha_G<1/4$.
Here, given BH and SF masses satisfying this inequality, %such that the same condition is satisfied, 
we obtain a family of self-gravitating solutions parameterized by the SF amplitude $A$ at the horizon. Although the solutions depend on the three
parameters $M_{BH}$, $m_\phi$, and $A$, configurations with the same $\alpha_G$ and $A$ are related to each other by a simple re-scaling.
This facilitates considering a wide set of astrophysical realizations (see Table~\ref{Tab} for relevant examples).
%\st{The study of ultralight DM + supermassive BH systems is a long-standing problem}
Among these realizations, the ultralight DM + supermassive BH system is a highly relevant one.
%~\cite{Barranco:2011eyw, Barranco:2012qs,Barranco:2013rua,Barranco:2017aes,Urena-Lopez:2002nup,Cruz-Osorio:2010nua,Davies:2019wgi,Bar:2019pnz,Hui:2019aqm,Clough:2019jpm,Li:2020qva,Cardoso:2022nzc,Pantig:2022sjb,Aurrekoetxea:2023jwk,DellaMonica:2023dcw,DeLuca:2023laa,Berezhiani:2023vlo,Aurrekoetxea:2024cqd}.
A key challenge in these systems arises from the vastly different scales between the BH and the DM halo core,
%that is described by a classical scalar field excitation~\cite{Hui:2021tkt}, 
resulting in configurations with $\alpha_G \sim 10^{-6}$, $M_T / M_BH \sim 10^3$, and $R_{\textrm{Sch}}/R_{\textrm{core}}\sim 10^{-10}$. Whereas we were not able to reach values below $\alpha_G\sim 0.1$ using the approach in~\cite{Barranco:2017aes}, here, we show that combining the use of horizon-penetrating coordinates with  adaptive grids allows one to construct configurations with %\st{$\alpha_G \sim 10^{-6}$ and $M_T/M_{BH}\sim 10^6$} 
relevant galactic parameters which last for time scales of the order of the Hubble time. Furthermore, we can follow very precisely
%\juan{In our previous work \cite{Barranco:2017aes} we performed the first approach to describe a self-gravitating scalar field in the presence of a spherically symmetric black hole. Nevertheless, in that work we constructed the solutions in terms of Schwarzschild coordinates that have a divergence at the horizon and additionally, we were unable to construct solutions with $\alpha_G \sim 10^{-6}$. In this work we avoid the singulartity at the horizon with a new set of variables ... explicar and thus we can reach trhe horizon and study the distributrion of scalar field and discuss the spike or no-.spike explicar.... Furthermore, because a new code with addaptative grid we have reached configurations with $\alpha_G=3.2\times 10^{-6}$ contrary to our previous work where the configurations were obtained for $\alpha_G \sim 0.1$. Thus we wre able to reach astrophyiscally connfigurations with time scales of the order of Hubble and wih $M_T/B_{BH} \sim 10^{6}$ etc...}
the self-gravitating SF's behavior from scales many orders of magnitude larger than the BH radius all the way to the apparent horizon, and explore the energy density in the BH's vicinity even in the strong gravity regime.
This allows us, for the first time, to study the existence and characteristics of
SFDM spikes~\cite{Gondolo:1999ef,Sadeghian:2013laa}.

%%%%%%%%%%%%%%%%%%%%%%%%%%%%%%%
% The quasi-stationary model
%%%%%%%%%%%%%%%%%%%%%%%%%%%%%%%

%{\it Quasi-stationary model.}|
\section{Quasi-stationary model}
We consider a complex SF $\phi$ of mass $m_\phi$ minimally coupled to gravity.\footnote{Current observations strongly constrain the possibility of a non-gravitational interaction between dark matter and the standard model of particle physics. We also neglect interactions with other fields through gravity, which would indeed be interesting to study but lies beyond the scope of this article. For ultralight axions, self-couplings do arise, but they typically scale with the axion mass, which
is very small in these models.}
A key feature that allows us to describe our gravitational atoms accurately is the use of horizon-penetrating coordinates (see also~\cite{2014CQGra..31s5006C,Gregory:2018ghc} for related work in the context of BHs in cosmology). Specifically, we represent the spherically symmetric spacetime metric as
\begin{equation}
ds^2 = -a^2 dt^2 + dr^2 + \frac{2m}{r}(a dt + dr)^2 
 + r^2 d\Omega^2,
\label{Eq:Metric}
\end{equation}
where $a$ and $m$ are positive functions of the time and areal radial coordinates $(t,r)$, and $d\Omega^2$ is the usual line element on the unit two-sphere. In this section only, we use Planck units such that $c=\hbar=G=1$. When $a=1$ and $m$ is constant Eq.~(\ref{Eq:Metric}) reduces to the Schwarzschild metric in the Kerr-Schild form, and the coordinates $(t,r)$ are regular on the future horizon $r=2m$ \cite{Chandrasekhar:1985kt}. In general, the surface $r = 2m$ in the spacetime described by the metric~(\ref{Eq:Metric}) is a non-expanding horizon~\cite{AshtekarBadri-LivRev}, along which the expansion with respect to the outward null vector $(\nabla^\mu r)\partial_\mu = a^{-1}\partial_t$ vanishes. 

For the following, we focus on quasi-stationary solutions (see~\cite{Barranco:2017aes} for more details on this approximation in terms of Schwarzschild-like coordinates), for which the SF and its derivative $D_0\phi$ along the future-directed unit normal to the $t=const$ time slices have the form 
\begin{equation}
\phi(t,r) = e^{st}\psi(r),\qquad
D_0\phi(t,r) = e^{st}\theta(r),
\label{Eq:Scalar}
\end{equation}
with $s = \sigma + i\omega$, where $\sigma < 0$ is the decay rate and $\omega$ the oscillation frequency. Substituting Eqs.~(\ref{Eq:Metric},\ref{Eq:Scalar}) into the EKG equations $G_{\mu\nu}=\nabla_\mu\phi^*\nabla_\nu\phi + \nabla_\nu\phi^*\nabla_\mu\phi - g_{\mu \nu}(\nabla_{\alpha}\phi^*\nabla^{\alpha}\phi + m_{\phi}^2\phi^*\phi)$ and $\nabla_\mu\nabla^\mu\phi-m_{\phi}^2\phi=0$, and defining \mbox{$\gamma:=\sqrt{1 + 2m/r}$} yields the system
\begin{subequations}\label{Eq.system}
\begin{eqnarray}
&&\left(1 - \frac{2m}{r} \right)\psi'' = B\psi' + C\psi,
\label{Eq:Psi}\\
&&2m' = e^{2\sigma t} r^2\left[ \gamma^2 \left| \frac{s}{a}\psi \right|^2 
 + m_\phi^2 |\psi|^2 + \left(1 - \frac{2m}{r} \right) \left| \psi' \right|^2 \right],
\qquad
\label{Eq:m}\\
&&\frac{a'}{a} = e^{2\sigma t} r \left| \frac{s}{a}\psi - \psi' \right|^2,
\label{Eq:a}
\end{eqnarray}
\end{subequations}
where a prime denotes derivation with respect to $r$ and %where
\begin{subequations}\label{Eq:BC}
\begin{eqnarray}
B := -\frac{2}{r}\left( 1 - \frac{m}{r} \right) - \frac{4m}{r}\frac{s}{a}
%\nonumber\\
 + \frac{e^{2\sigma t} r}{\gamma^2}\biggl[ 
  (\gamma^4-1)\left| \frac{s}{a}\psi \right|^2 + \gamma^2 m_\phi^2 |\psi|^2 %\nonumber\\
   + (1+\gamma^2)\left( 1 - \frac{2m}{r} \right)\re\left( \frac{s^*}{a}\psi^*\psi' \right)
  \biggr],
\label{Eq:B} \\
%\end{eqnarray}
%and
%\begin{eqnarray}
C := m_\phi^2 + \gamma^2\left( \frac{s}{a} \right)^2 - \frac{2m}{r^2}\frac{s}{a}
%\nonumber\\
 - e^{2\sigma t} r\biggl[ \left| \frac{s}{a}\psi \right|^2 + m_\phi^2 |\psi|^2
   + \left( 1 - \frac{2m}{r} \right)\left| \psi' \right|^2%\nonumber\\
   - \left( 1 - \frac{2m}{r} \right)\re\left( \frac{s^*}{a}\psi^*\psi' \right)
\biggr] \frac{s}{a}. 
\label{Eq:C}
\end{eqnarray}
\end{subequations}
Furthermore, one obtains
\begin{equation}
\dot{m} = r^2 a e^{2\sigma t}\left[ \frac{2m}{r}\left| \frac{s}{a}\psi \right|^2
 + \left( 1 - \frac{2m}{r} \right)\re\left( \frac{s^*}{a}\psi^*\psi' \right) \right]
\label{Eq:mdot}
\end{equation}
for the time derivative $\dot{m}$ of the mass function, which allows one to estimate the mass accretion rate $\dot{M}_{BH} := \left. \dot{m} \right|_{r=2m}$ at the horizon.
%\st{Einstein's equations also yield an equation for the time derivative of $m$ which will not be used in our model}. 
The quasi-stationary approximation consists in replacing the factors $e^{2\sigma t}$ in Eqs.~(\ref{Eq:m},\ref{Eq:a},\ref{Eq:BC},\ref{Eq:mdot}) by $1$ 
%\st{, such that the metric coefficients $a$ and $m$ can be assumed to be time-independent.}
and in assuming that the metric coefficients $a$ and $m$ are time-independent.
After the specification of appropriate boundary conditions at $r=2m$ and $r\to\infty$, the resulting system~(\ref{Eq.system}) for $(\psi,m,a)$ constitutes a nonlinear eigenvalue problem for the complex frequency $s$. Their corresponding solutions $(\phi,m,a)$, with $\phi$ given by Eq.~(\ref{Eq:Scalar}), yield approximate solutions of the EKG equations which are expected to be accurate on time scales $t\ll  \min\{ t_0,t_{\rm accr} \}$, where $t_0:=1/|\sigma|$ and $t_{\rm accr} := M_{BH}/\dot{M}_{BH}$. Note also that these fields provide \emph{exact} solutions of the initial data constraints, as follows from the arguments at the end of Section~II in~\cite{Barranco:2017aes}.

At $r\to \infty$ we demand $\psi\to 0$, as required by asymptotic flatness. At $r=2m$ a difficulty arises because the coefficient in front of $\psi''$ in Eq.~(\ref{Eq:Psi}) vanishes there. 
To treat this problem, we write 
\begin{equation}
r = r_0(1 + z),\quad
2m = r_0\left[ 1 + z p(z) \right], \label{eq:m-p}
\end{equation}
where $r_0 = 2m(r_0) := 2M_{BH} > 0$ is a given parameter which is equal to the apparent horizon radius and $p$ is a positive function of the dimensionless coordinate $z$. Since $1 - \frac{2m}{r} = \frac{z}{1+z}(1-p)$ we require that $p < 1$ for all $z\geq 0$ in order to avoid the presence of a second horizon in the region $r > r_0$. In terms of the new variables $z$ and $p$, Eq.~(\ref{Eq:Psi}) can be brought into the form
\begin{equation}
z\frac{d^2\psi}{dz^2} + \Delta(z)\frac{d\psi}{dz} = \Gamma(z)\psi,
\label{Eq:PsiInTermsOfz}
\end{equation}
with the coefficients $\Delta(z) := -\frac{rB}{1-p}$ and 
$\Gamma(z) := \frac{r_0 r C}{1-p}$. If a solution that is regular at $z=0$ exists, then evaluating Eq.~(\ref{Eq:PsiInTermsOfz}) at $z=0$ yields
\begin{eqnarray}
\left. \frac{1}{\psi}\frac{d\psi}{dz} \right|_{z=0}
 &=& \frac{\Gamma(0)}{\Delta(0)},
\label{Eq:PsiFirstDeriv}
\end{eqnarray}
which allows one to explicitly determine the first derivative of $\psi$ at $z=0$. Assuming without loss of generality that $a(r_0) = 1$ and noticing that 
$p(0) 
 = \left. 2m' \right|_{r=r_0}$
can be evaluated using Eq.~(\ref{Eq:m}), one finds
\begin{subequations}
\begin{eqnarray}
p(0) &=& 2\left( 2\alpha_G^2 + \left| \bar{s} \right|^2 \right) |A|^2,
\label{Eq:p0}\\
\Delta(0) &=& 1 + \frac{4\bar{s} + | \bar{s} |^2 |A|^2}{2(1-p(0))},
\label{Eq:Delta0}\\
\Gamma(0) &=& \frac{4\alpha_G^2 + 2\bar{s}^2 - \bar{s} - \left( p(0) - \left| \bar{s} \right|^2 |A|^2 \right)\bar{s}}{1-p(0)},
\label{Eq:Gamma0}
\end{eqnarray}
\end{subequations}
where we have used $r_0 m_\phi = 2M_{BH} m_\phi = 2\alpha_G$, and defined $\bar{s} := r_0 s$ and $A := \psi(r_0)$.

Summarizing, the apparent horizon boundary conditions are
\begin{equation}
\psi(r_0) = A, \;\;
\psi'(r_0) = \frac{A}{r_0}\frac{\Gamma(0)}{\Delta(0)},\;\;
2m(r_0) = r_0,\;\;
a(r_0) = 1 
\label{Eq:AHBC},
\end{equation}
where $\Delta(0)$ and $\Gamma(0)$ are given in Eqs.~(\ref{Eq:Delta0},\ref{Eq:Gamma0}). Besides $r_0 = 2M_{BH}> 0$, the only other free parameter at the horizon is the SF's amplitude $A$ whose norm should be chosen small enough such that $p(0) < 1$. We have assumed that $a(r_0) = 1$ to simplify the calculations; however one can recover the general case by replacing $s$ with $s/a(r_0)$ in the expressions above. More precise horizon boundary conditions which include an explicit expression for $\psi''(r_0)$ are provided in the Appendix.

%%%%%%%%%%%%%%%%%%%%%%%%%%%%%%%
%\section{Results}
%%%%%%%%%%%%%%%%%%%%%%%%%%%%%%%

%{\it Results.}|
\section{Results}
\begin{figure*}[htb]
    \begin{subfigure}[b]{0.49\textwidth}
    \includegraphics[width=\textwidth]{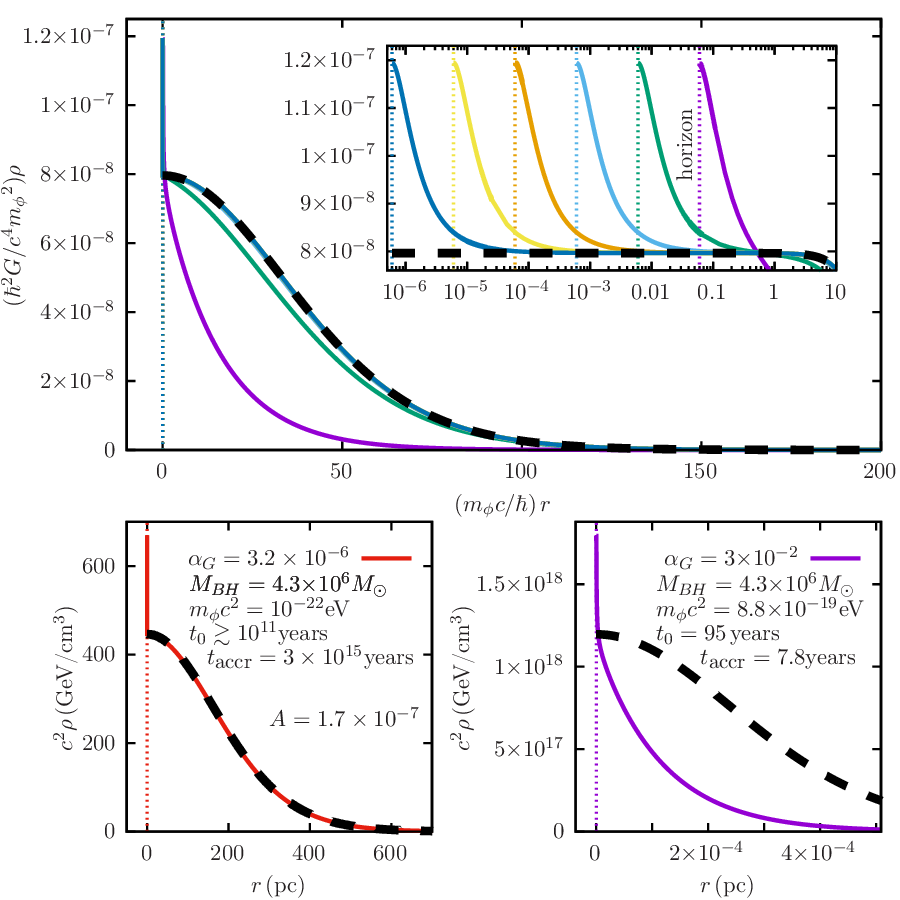}
    \caption{Energy density}
    \end{subfigure}
    \begin{subfigure}[b]{0.49\textwidth}
    \includegraphics[width=\textwidth]{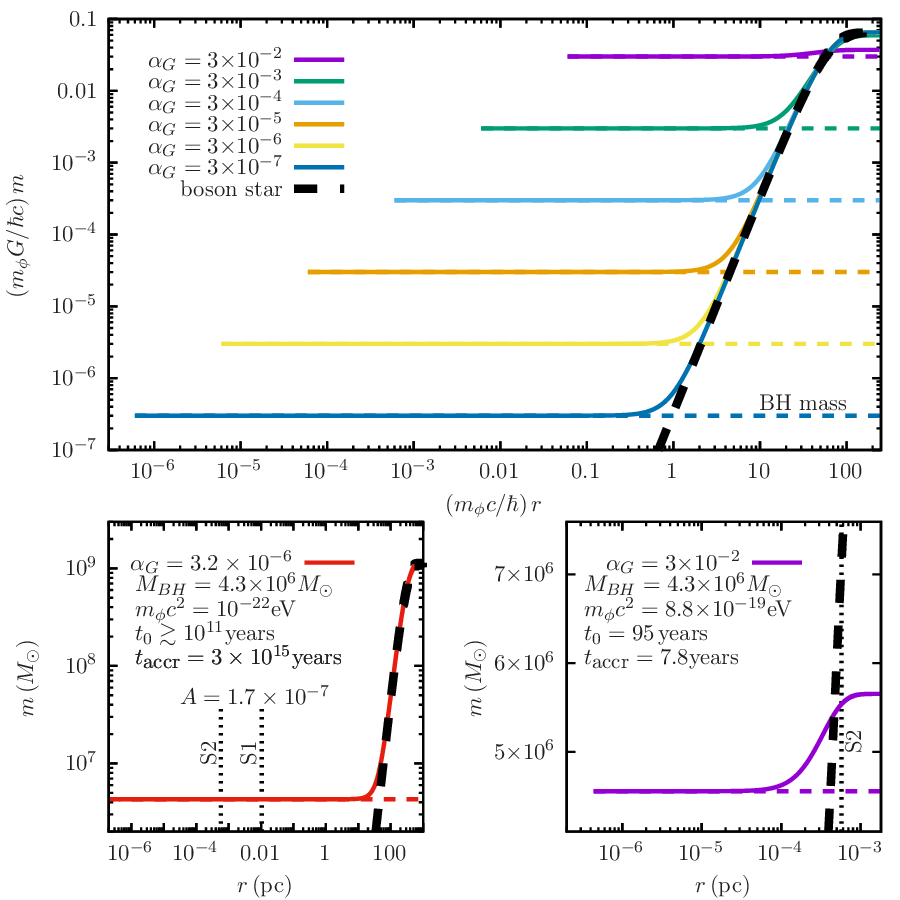}
    \caption{Mass function}
    \end{subfigure}
    \caption{ \justifying
        Energy density and mass function at $t=0$ for solutions with various values of $\alpha_G$, together with those for BSs with the same central amplitude in each case. The amplitude is $A=10^{-3}$ in all cases except the one indicated as $A=1.7\times 10^{-7}$.
        Color dotted lines in the density plots indicate each apparent horizon's location, color dashed lines in the mass plots indicate each BH mass, and black dotted lines indicate the pericenter distance of stars S1 and S2 in the case of Sgr~A*. The curves belonging to $\alpha_G$ from $3{\times}10^{-4}$ to $3{\times}10^{-7}$ in the first plot are not clearly seen in most of the region because they are superposed almost perfectly with the BS curve.
    \label{f:spikecomp}}
\end{figure*}

The solutions are obtained by numerical integration of Eqs.~(\ref{Eq.system}) using a shooting algorithm, with boundary conditions at the horizon obtained from Eq.~(\ref{Eq:AHBC}), and demanding $\psi\rightarrow0$ exponentially for large $r$. The numerical algorithm is based on the one described in~\cite{Press1992Numerical}, but using the adaptive step solver presented in~\cite{Radhakrishnan1993}, which provides high accuracy at low computational cost. 
We concentrate on the ``ground-state'' 
solutions (i.e. those with no zeros of $|\psi(r)|$) and leave the study of the excited states for future work. 
Some examples are shown in Fig.~\ref{f:spikecomp}, where the energy density $\rho := m'/(4\pi r^2)$ and mass function $m$ at $t=0$ are shown for different parameters. 

Let us concentrate first on the top plots in Fig.~\ref{f:spikecomp}, in which all configurations have $A=10^{-3}$. For small enough $\alpha_G$, the solutions (re-scaled by $m_{\phi}$) look very similar to each other and to a boson star (BS) of the same amplitude %for large scales, 
almost everywhere, except for a very small region close to the horizon, showing a wide core. As $\alpha_G$ becomes larger, the solutions begin to differ more and more, displaying a sharper profile. 
Although it is possible to see an over-density (or spike)~\cite{Gondolo:1999ef,DellaMonica:2023dcw,Davies:2019wgi,Li:2020qva,Pantig:2022sjb} near the horizon,  
its contribution to the mass function is negligible in most cases, even close to the apparent horizon, where $m$ is practically identical to the BH mass. 
Note that, for fixed $A$, solutions with different values of $\alpha_G$ have density spikes with approximately the same height. 
This is consistent with the fact that at the horizon $8\pi\rho = r_0^2 p_0 = (m_\phi^2 + 2|s|^2|)|A|^2$, where we have used Eq.~(\ref{Eq:p0}), and that $s\approx m_\phi$ in these cases.

To exemplify with a particular physical system, we concentrate on Sgr~A* setting $M_{BH}=4.3{\times}10^{6} M_{\odot}$, and consider two 
cases, shown in the bottom plots of Fig.~\ref{f:spikecomp}, (i) $\alpha_G=3.2{\times}10^{-6}$ ($m_{\phi}c^2=10^{-22}\unit{eV}$), $A=1.7\times 10^{-7}$ and (ii) $\alpha_G=3{\times}10^{-2}$ ($m_{\phi}c^2=8.8{\times}10^{-19}\unit{eV}$), $A=10^{-3}$.  
We see that the configuration in case (i) 
is almost identical to a BS with the same parameters,  
having enough mass to explain the dispersion velocities of bulge stars in spiral galaxies, 
where supermassive BHs inhabit but play an insignificant role in the galaxy dynamics~\cite{DeMartino:2018zkx}.
S-stars' trajectories and the BH's image would be unaffected in this case. 
On the other hand, in case (ii) the configuration departs strongly from that of a BS. 
In this case, the SF has a high concentration around the BH. 
Thus, the contribution of the self-gravitating SF mass in that region is important, and it could even affect the dynamics of S-stars. 
Note also the huge difference in the characteristic times $t_0$ and $t_{\rm accr}$ of these configurations. 
We have $t_0\gtrsim 10^{11}\, \unit{years}$~\footnote{In this case $\sigma$ is many orders of magnitude smaller than $\omega$, which makes it difficult to obtain its value accurately. 
We are still able to obtain an approximate upper bound, and hence an approximate lower bound for $t_0$.} and $t_{\rm accr}=3\times 10^{15}\, \unit{years}$ in case (i), consistent with galactic lifetimes, and the extremely small values $t_0=95\, \unit{years}$ and $t_{\rm accr}=7.8\, \unit{years}$  in case (ii).

\begin{figure}[htb]
\centerline{\includegraphics[width=1\textwidth]{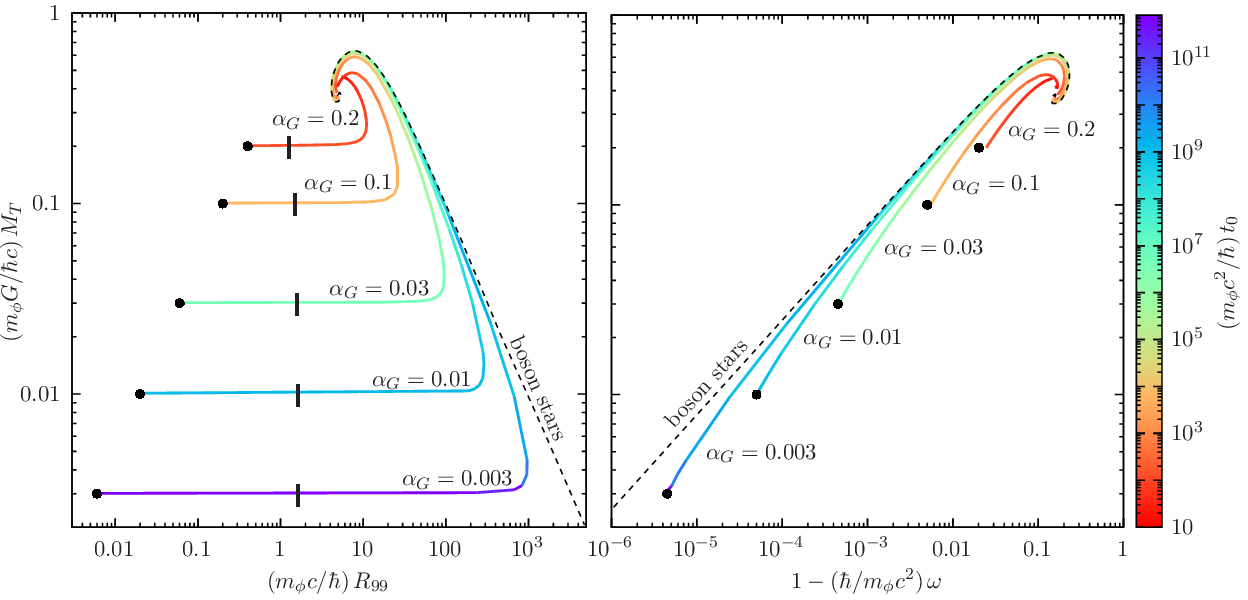}}
    \caption{ \justifying
    $M_T$ vs $R_{99}$ (left panel)  and $M_T$ vs $\omega$ (right panel) of solutions with various values of $\alpha_G$, varying $A$. 
    The color map indicates the characteristic time $t_0$. 
    Black dots indicate, for each $\alpha_G$, the points where $M_T=M_{BH}$, $R_{99}=R_{\rm Sch}$ and $(\hbar/m_{\phi}c^2)\omega_n=\sqrt{1-\alpha_G^2/n^2}$ (with $n=1$).
    Short vertical lines indicate the locations of the test field's effective potential minima.
    \label{f:MR99}} 
\end{figure}
Some solutions' properties throughout the parameter space are shown in Fig.~\ref{f:MR99}, where the total mass $M_T:= m(r\rightarrow\infty)$ is shown as a function of the radius $R_{99}$~\footnote{As usual, we define the radius $R_{99}$ as that of a sphere containing $99\%$ of the total mass.} and the frequency $\omega$. 
Each curve is obtained by fixing $\alpha_G$ and varying $A$.
For small $A$, the test field limit is recovered, at which point the solutions accumulate as $A\rightarrow 0$.
The black dots indicate the values $M_T=M_{BH}$, $R_{99}=R_{\rm Sch}$ and $(\hbar/m_{\phi}c^2)\omega_n=\sqrt{1-\alpha_G^2/n^2}$ (with $n=1$). 
%
%As expected, the first two values are recovered for small $A$, whereas the last one is recovered when both $A$ and $\alpha_G$ are small enough~\cite{Barranco:2013rua}.
%
As expected, the values $M_T=M_{BH}$ and $R_{99}=R_{\rm Sch}$ are recovered for small $A$, whereas $(\hbar/m_{\phi}c^2)\omega_n=\sqrt{1-\alpha_G^2/n^2}$ is recovered when both $A$ and $\alpha_G$ are small enough~\cite{Barranco:2013rua}.
In another region of the parameter space, the properties shown become identical to those of BSs, indicated by a dashed line. Note that although this limit occurs at large $A$, it is not necessarily true that it is obtained from $A\rightarrow \infty$ at fixed $\alpha_G$.
We also see that the characteristic times $t_0$, indicated by the color map, span many orders of magnitude.

The condition $\alpha_G< 1/4$ for the existence of quasi-stationary solutions in the test field limit is recovered empirically for the self-gravitating case, as we were able to find solutions close to, but not beyond that limit.
Additionally, for all the solutions obtained in this work, $M_T$ is bounded from above by the same value as BSs, that is $(m_{\phi} G/\hbar c) M_T<0.633$, with solutions close to this maximum for large enough $A$. 
In the opposite limit, we have $M_T>M_{BH}$, and solutions get close to that minimum for small $A$. 
Regarding the characteristic time $t_0$, we see that it is limited from above by that of the test filed limit for each value of $\alpha_G$. 
This limit can be evaluated in the small $\alpha_G$ approximation as shown in~\cite{Barranco:2012qs}.
Finally, we provide size estimates. 
The minimum size in each case is simply given by the Schwarzschild radius, $R_{99}^{\rm min}=R_{\rm Sch}$. 
Lacking a simple expression for the maximum size, we can provide bounds. 
An upper bound, $R_{99}^{{\rm max}+}$, is given by the BSs curve, as is evident from Fig.~\ref{f:MR99}.
Using the approximation $M_T = 10\, \hbar^2/(G m_{\phi}^2 R_{99})$, valid in the non-relativistic limit, we get $R_{99}^{{\rm max}+} = 10\, \hbar^2/(G m_\phi^2 M_{BH})$.
To estimate a lower bound, $R_{99}^{{\rm max}-}$, we note that the maximum radius is to the right of the test field region. 
Hence, a lower bound is given by the location of the test field effective potential minimum~\cite{Barranco:2013rua}. 
Although this is not a very tight bound, it can still be useful for evaluating certain astrophysical scenarios' plausibility.

\setlength{\tabcolsep}{4pt}
\begin{table*}[htb]
\caption{ \justifying
    Limits of $M_T$, $R_{99}$ and $t_0$ for combinations of $m_{\phi}$ and $M_{BH}$ motivated by various astrophysical scenarios. 
    Note that for $\alpha_G>1/4$ there are no solutions, hence those cases are not included. 
    For clarity, we only show orders of magnitude.
\label{Tab}}
\begin{tabular}{l  l  l  l  l  l  l  l  l  l}
\hline 
\hline
$\alpha_G$ & $m_\phi c^2$ & $M_{BH}=$                    &  $M_{T}^{\rm max}$ & $R_{99}^{\rm min}=$ & $R_{99}^{{\rm max}-}$ & $R_{99}^{{\rm max}+}$ & $t_0^{\rm max}$ & Motivation   & Motivation         \\ 
           & (eV)         & $M_{T}^{\rm min}\,(M_\odot)$ &  $(M_\odot)$       & $R_{\rm Sch}$ (pc)  & (pc)                  & (pc)                  & (years)         & for $m_\phi$ & for $M_{BH}$       \\
\hline
$10^{-29}$ & $10^{-22}$   & \;$10^{-17}$                 &  $10^{12}$         & \;$10^{-30}$        & $0.1$                 & $10^{29}$             & $10^{144}$      & Fuzzy DM     & Primordial BH      \\
$10^{-25}$ & $10^{-18}$   & \;$10^{-17}$                 &  $10^{8}$          & \;$10^{-30}$        & $10^{-5}$             & $10^{21}$             & $10^{120}$      & ALP          & Primordial BH      \\
$10^{-25}$ & $10^{-22}$   & \;$10^{-13}$                 &  $10^{12}$         & \;$10^{-26}$        & $0.1$                 & $10^{25}$             & $10^{124}$      & Fuzzy DM     & Microlensing limit \\
$10^{-21}$ & $10^{-18}$   & \;$10^{-13}$                 &  $10^{8}$          & \;$10^{-26}$        & $10^{-5}$             & $10^{17}$             & $10^{100}$      & ALP          & Microlensing limit \\
$10^{-12}$ & $10^{-5}$    & \;$10^{-17}$                 &  $10^{-5}$         & \;$10^{-30}$        & $10^{-18}$            & $10^{-5}$             & $10^{42}$       & QCD axion    & Primordial BH      \\
$10^{-12}$ & $10^{-22}$   & \;$1$                        &  $10^{12}$         & \;$10^{-13}$        & $0.1$                 & $10^{12}$             & $10^{59}$       & Fuzzy DM     & Stellar BH         \\
$10^{-10}$ & $10^{-22}$   & \;$100$                      &  $10^{12}$         & \;$10^{-11}$        & $0.1$                 & $10^{10}$             & $10^{49}$       & Fuzzy DM     & LIGO detection     \\
$10^{-8}$  & $10^{-5}$    & \;$10^{-13}$                 &  $10^{-5}$         & \;$10^{-26}$        & $10^{-18}$            & $10^{-9}$             & $10^{22}$       & QCD axion    & Microlensing limit \\
$10^{-8}$  & $10^{-18}$   & \;$1$                        &  $10^{8}$          & \;$10^{-13}$        & $10^{-5}$             & $10^{4}$              & $10^{35}$       & ALP          & Stellar BH         \\
$10^{-6}$  & $10^{-18}$   & \;$100$                      &  $10^{8}$          & \;$10^{-30}$        & $10^{-5}$             & $100$                 & $10^{25}$       & ALP          & LIGO detection     \\
$10^{-6}$  & $10^{-22}$   & \;$10^{6}$                   &  $10^{12}$         & \;$10^{-7}$         & $0.1$                 & $10^{6}$              & $10^{29}$       & Fuzzy DM     & Sgr~A*             \\
$10^{-3}$  & $10^{-22}$   & \;$10^{9}$                   &  $10^{12}$         & \;$10^{-4}$         & $0.1$                 & $10^{3}$              & $10^{14}$       & Fuzzy DM     & M87*               \\
$10^{-2}$  & $10^{-18}$   & \;$10^{6}$                   &  $10^{8}$          & \;$10^{-7}$         & $10^{-5}$             & $10^{-2}$             & $10^{5}$        & ALP DM       & Sgr~A*             \\
$10^{-1}$  & $10^6$       & \;$10^{-17}$                 &  $10^{-16}$        & \;$10^{-30}$        & $10^{-29}$            & $10^{-27}$            & $10^{-24}$      & MeV DM       & Primordial BH      \\
\hline
\hline
\end{tabular}
\end{table*} 

We now consider some relevant astrophysical examples corresponding to particular values of $M_{BH}$ and $m_{\phi}$ (and hence $\alpha_G$), shown in Table~\ref{Tab}. 
The values chosen for $M_{BH}$ are motivated by primordial BHs as dark matter~\cite{Bicknell1976,Carr:2020xqk}, limit BH by micro-lensing~\cite{Niikura:2017zjd}, stellar BHs, LIGO BH detection~\cite{LIGOScientific:2016aoc}, Sgr~A*~\cite{Schodel:2002py,Ghez:2008ms} and M87*; and those of $m_{\phi}$ by fuzzy DM~\cite{Hu:2000ke,Matos:1999et,Matos:2000ss}, axion-like particles DM~\cite{Marsh:2015xka}, QCD axion DM and MeV DM. 
Each combination of these sets fixes the two masses, leaving a family of solutions parameterized by $A$ whenever $\alpha_G<1/4$, and no solutions otherwise. 
For the cases with solutions, we show limiting values of $M_T$, $R_{99}$, and $t_0$ in the table. %, and indicate, based on these values, whether each scenario might be plausible in our model. 

%%%%%%%%%%%%%%%%%%%%%%%%%%%%%%%
%  Discussion and Outlook
%%%%%%%%%%%%%%%%%%%%%%%%%%%%%%%

%{\it Discussion}.|
\section{Discussion}
The configurations obtained here resemble the quasi-bound states for small values of $A$, while they depart from the hydrogen-like bosonic clouds and yield new self-gravitating configurations for large $A$. In a suitable limit, these configurations approach BSs.
Although we have not yet studied their stability with respect to small perturbations, there are some indications that allow us to conjecture that our configurations may be stable and generic: first, our previous work in~\cite{Barranco:2012qs,Barranco:2013rua} shows that in the test-field limit these configurations arise from generic initial data. Second, in~\cite{Barranco:2017aes}, numerical evolutions of the EKG system in spherical symmetry starting from initial data representing the quasi-stationary solution correctly reproduce the oscillation frequencies and decay rates for configurations with $\alpha_G\sim 0.1$ and $M_{T}/M_{BH}\sim 1.5$. Finally, it is known that ground state BS possess a stable branch which is the one connecting with the Newtonian configurations~\cite{Liebling:2012fv}.

Most configurations display a sort of {\it decoupling}, behaving quite differently at different scales. 
Typically, the relatively small region near the BH (which can however be much larger than the BH itself) has a mass function practically identical to that of the BH alone, whereas at large scale (usually on an extension many orders of magnitude the BH radius) the mass function coincides with that of a BS (without any effect of the central BH). 

Of particular interest are cases where the SF is ultralight and the BH is supermassive, for which our solutions are consistent with long-lived DM halo cores with BHs in their center; see examples in Table~\ref{Tab}. 
In particular, the characteristic times of these configurations are larger than cosmological times.
In those cases, a density spike is present very close to the apparent horizon.
However, its contribution to the mass function is negligible compared to that of the BH, and we do not expect it would alter nearby trajectories significantly. 
This result contrasts with that reported for cold DM in~\cite{Gondolo:1999ef}.

%%%%%%%%%%%%%%%%%%%%%%%%%%%
%%%   ACKNOWLEDGMENTS   %%%
%%%%%%%%%%%%%%%%%%%%%%%%%%%

\section*{Acknowledgments}
%{\it Acknowledgments.}|
This work was partially supported by the CONAHCyT Network Projects No.~376127 ``Sombras, lentes y ondas gravitatorias generadas por objetos compactos astrof\'\i sicos" and No.~304001 ``Estudio de campos escalares con aplicaciones en cosmología y astrofísica", by CONAHCyT-SNII,
by  PAPIIT-UNAM projects IN100523 and IN110523, by CIC grant No.~18315 of the Universidad Michoacana de San Nicol\'as de Hidalgo,
by the Center for Research and Development in Mathematics and Applications (CIDMA) through the Portuguese Foundation for Science and Technology (FCT - Fundação para a Ciência e a Tecnologia), references UIDB/04106/2020 and UIDP/04106/2020, and by the programme HORIZON-MSCA2021-SE-01 Grant No.~NewFun-FiCO101086251. A.D.T. acknowledges support from
DAIP project CIIC 2024 198/2024.  DN acknowledges the sabbatical support given by PAPIIT-UNAM
%the Programa de Apoyos para la Superaci\'on del Personal Acad\'emico de la Direcci\'on General de Asuntos del Personal Acad\'emico (PAPIIT) de la Universidad Nacional Aut\'onoma de M\'exico (UNAM) 
in the elaboration of the present work. JB, AB, DN and OS thank the Mathematics Department at the University of Aveiro for their hospitality, where part of this work was completed.
MM acknowledges financial support from CONICET (PIP 11220210100914CO).

\appendix
\section*{Appendix}

Here, we further develop the apparent horizon boundary conditions and derive an explicit expression for $\psi''(r_0)$. This allows one to start the integration of Eqs.~(5) at $r=r_0$ and even continue the solution in a small region inside the horizon, which should be useful for numerical evolutions.

Given the central amplitude $A = \psi(r_0)$ of the field, evaluating Eq.~(9) and its first derivative at $z=0$ yields
\begin{eqnarray}
\psi'(r_0) &=& \frac{A}{r_0}\frac{\Gamma_0}{\Delta_0},\\
\psi''(r_0) &=& \frac{A}{r_0^2}\frac{\Gamma_0^2 - \Gamma_0\Delta_1 + \Gamma_1\Delta_0}{\Delta_0(1+\Delta_0)},
\label{Eq:ddPsi}
\end{eqnarray}
with $\Gamma_k$ and $\Delta_k$ denoting the $k$th order coefficients of
\begin{equation}
\Delta(z) := -\frac{rB}{1-p},\qquad
\Gamma(z) := \frac{r_0 r C}{1-p}
\label{eqs:DG}
\end{equation}
in their Taylor expansion around $z=0$. For the following, we compute these coefficients for $k=0,1$, as needed for Eq.~(\ref{Eq:ddPsi}). 

From Eq.~\eqref{eqs:DG}, we first obtain
\begin{eqnarray}
&& \Delta_0 = -\frac{r_0 B_0}{1-p_0},\qquad
\Gamma_0 = -\frac{r_0^2 C_0}{1-p_0},\\
&& \Delta_1 = -\frac{r_0}{1-p_0}\left( B_1 + \frac{1-p_0 + p_1}{1-p_0}\,B_0 \right),\\
&& \Gamma_1 = \frac{r_0^2}{1-p_0}\left( C_1 + \frac{1-p_0 + p_1}{1-p_0}\,C_0 \right),
\end{eqnarray}
where $p_k$, $B_k$ and $C_k$ refer again to the Taylor coefficients at $z=0$. Evaluating Eqs.~(6a,6b) at $z=0$ one first finds
\begin{align}
r_0 B_0 &= -\left[ 1 + 2\bar{s} - \left( 8{\alpha_G}^2 + 3|\bar{s}|^2\right)\frac{|A|^2}{2}\right], \label{eq:B0}
\\
r_0^2 C_0 &= 4{\alpha_G}^2 + 2|\bar{s}|^2 - \bar{s}\,\left[ 1 + \left( 4{\alpha_G}^2 + |\bar{s}|^2\right)|A|^2\right], \label{eq:C0}
\end{align}
where we recall the definitions $\bar{s} := r_0 s$ and $\alpha_G = r_0 m_\phi/2$. Together with the expression for $p_0$ in Eq.~(11a) this yields the results for $\Delta_0$ and $\Gamma_0$ listed in Eqs.~(11b,11c).

Next, in order to compute $p_1$, we differentiate Eq.~(8) twice and evaluate at $z=0$, which yields $p_1 = r_0\left. m'' \right|_{r=r_0}$. To evaluate the second derivative of $m$, we use Eq.~(5b) together with the following identities which can be inferred from Eqs.~(8,5c) and the definition of $\gamma$:
\begin{equation}
%\left. m' \right|_{r=r_0} = \frac{p_0}{2},\quad
\left. (\gamma^2)' \right|_{r=r_0} = \frac{p_0-1}{r_0},\qquad
\left. a' \right|_{r=r_0} = \frac{1}{r_0}\left| \bar{s} - T_0 \right|^2|A|^2,
\label{Eq:Ids}
\end{equation}
where we have introduced $T_0 := \Gamma_0/\Delta_0$ to simplify the notation. This yields
\begin{eqnarray}
p_1 &=& \frac{p_0}{2}\left( 2 + T_0 + T_0^* \right)
 + \frac{1}{2}(p_0-1)\left( |\bar{s}|^2 - |T_0|^2\right)|A|^2
 - 2|\bar{s}|^2\left| \bar{s} - T_0 \right|^2 |A|^4.
\end{eqnarray}

Finally, differentiating Eqs.~(6a,6b) and using the identities~\eqref{Eq:Ids} again one obtains
\begin{eqnarray}
r_0 B_1 &=& h_a\,|T_0|^2 + h_b\,T_0 + h_c\, T_0^* + h_d,
\\
r_0^2 C_1 &=& j_a\,|T_0|^2 + j_b\,T_0 + j_c\, T_0^* + j_d, 
\end{eqnarray}
with coefficients
\begin{eqnarray}
h_a &=& \bar{s}\,|A|^2\,\left(2 - 3\,\bar{s}^*\,|A|^2\right), \nonumber \\
h_b &=& -\frac{|A|^2}{4}\,\left[6\,|A|^2\,\bar{s}^*\,\left(2\,{\alpha_G}^2 -|\bar{s}|^2\right) - 16\,{\alpha_G}^2 + 2\,|\bar{s}|^2 - 3\,\bar{s}^*\right], \nonumber \\
h_c &=& -\frac{|A|^2}{4}\,\left[6\,|A|^2\,\bar{s}\,\left(2\,{\alpha_G}^2 -|\bar{s}|^2\right) - 16\,{\alpha_G}^2 + 8\,\bar{s}^2  
 - 6\,|\bar{s}|^2- 3\,\bar{s} \right], \nonumber \\
h_d &=& \frac{|A|^4\,|\bar{s}|^2}{2}\,\left(10\,{\alpha_G}^2 - \,|\bar{s}|^2 \right) + 2\bar{s}
 + |A|^2\,\left[8\,{\alpha_G}^2\,\left(1-|\bar{s}|^2 \right) + |\bar{s}|^2\,\left(\frac{9}{4}-2\bar{s}\right) \right],
\nonumber \\
j_a &=& \bar{s}\,|A|^2\,\left[|A|^2\,\left(8\,{\alpha_G}^2 + 5\,|\bar{s}|^2\right) - 4\,\bar{s}^*\right],
\nonumber\\
j_b &=& -\frac{|A|^2\,\bar{s}}{2}\left[4\,|A|^2\,\bar{s}^*\,\left(3\,{\alpha_G}^2 + 2\,|\bar{s}|^2\right) +  2\,\left(4\,{\alpha_G}^2 + |\bar{s}|^2 - 4\,\left(\bar{s}^*\right)^2\right) + \bar{s}^*\right], \nonumber \\
j_c &=& -\frac{|A|^2\,\bar{s}}{2}\left[4\,|A|^2\,\bar{s}\,\left(3\,{\alpha_G}^2 + 2\,|\bar{s}|^2\right) + 8\,{\alpha_G}^2 - 6\,|\bar{s}|^2 + \bar{s}\right],  \nonumber \\
j_d &=& \bar{s}\,\left[|A|^4\,|\bar{s}|^2\,\left(4\,{\alpha_G}^2 + 3\,|\bar{s}|^2\right) + 2\,|A|^2\,\left(2\,{\alpha_G}^2\,\bar{s}^* - \bar{s}^*\,|\bar{s}|^2 - 4\,{\alpha_G}^2 - |\bar{s}|^2\right) -  \bar{s}^* + 2\right]. \nonumber
\end{eqnarray}
%

%%%%%%%%%%%%%%%%
%\subsection*{\label{sec:citeref} References}
%\bibliographystyle{unsrt}
%\bibliography{ref.bib}
%%%%%%%%%%%%%%%%

%%%%%%%%%%%%%%%%%%%%%%%%%%%%%%%

\end{document}